\documentclass[aps,amsfonts,amsmath,nofootinbib,showpacs,byrevtex,%
preprintnumbers,showkeys,preprint,prd]{revtex4} 
\usepackage{graphicx}
\usepackage{hyperref}
\usepackage{bm} 
\usepackage{tikz}
\usetikzlibrary{snakes}

\allowdisplaybreaks[1]
\def\be#1\ee{\begin{equation}#1\end{equation}}
\def\bal#1\eal{\begin{align}#1\end{align}}
\def\bat#1\eat{\begin{alignat}{2}#1\end{alignat}}
\def\bmu#1\emu{\begin{multline}#1\end{multline}}
\def\bga#1\ega{\begin{gather}#1\end{gather}}
\newcommand{\ba}{\begin{array}}
\newcommand{\ea}{\end{array}}
\newcommand{\n}{\notag}

\renewcommand{\d}{\partial}

\newcommand{\scs}{\scriptstyle}

\begin{document}

\title{Epsilon expansion for infrared Yang-Mills theory in Landau gauge}
\date{\today}

\author{Axel Weber}
\affiliation{Instituto de F\'isica y Matem\'aticas,
Universidad Michoacana de San Nicol\'as de Hidalgo,
Edificio C-3, Ciudad Universitaria, A. Postal 2-82, 58040 Morelia, 
Michoac\'an, Mexico}

\begin{abstract}
The study of the Dyson-Schwinger equations of Landau gauge
Yang-Mills theory has revealed two types of solutions for the 
gluon and ghost propagators, with a scaling and a massive 
(decoupling) behavior in the extreme infrared, respectively.
We show that both types of solutions are quantitatively reproduced by
applying renormalization group equations of Callan-Symanzik type in an
epsilon expansion to the infrared limit of Landau gauge
Yang-Mills theory when a mass term for the gluons is added to
the action. Only the decoupling solution corresponds to an 
infrared-stable fixed point in three and four space-time
dimensions and is hence expected to be physically
realized, in agreement with the results of recent lattice 
calculations.
\end{abstract}

\maketitle

The infrared (IR) behavior of gluon and ghost propagators in $SU(N)$
Yang-Mills theory in Landau gauge has been a subject of intense debate
over the last decade. The first results that have been obtained in the
extreme IR region, for momentum scales $p \ll \Lambda_{\text{QCD}}$,
came from the solution of an approximation to the full Dyson-Schwinger
(DS) equations in this theory \cite{SHA97}. They are nowadays known 
as scaling solutions, showing power-law behavior of the propagators in the IR.

For a quantitative description of these results, we introduce some notation.
The gluon and ghost propagators in $D$ Euclidean space-time dimensions are 
parameterized as
\bal
\langle A^a_\mu (p) A^b_\nu (-q) \rangle 
&= \frac{G (p^2)}{p^2} \left( \delta_{\mu \nu} - \frac{p_\mu p_\nu}{p^2}
\right) \delta^{ab} (2 \pi)^D \delta (p - q) \,, \n \\
\langle c^a (p) \bar{c}^b (-q) \rangle
&= \frac{F (p^2)}{p^2} \, \delta^{ab} (2 \pi)^D \delta (p - q) \,.
\label{FGdefs}
\eal
The scaling solutions show a power behavior of the dressing functions in 
the IR,
\be
G (p^2) \propto (p^2)^{- \alpha_G} \,, \qquad
F (p^2) \propto (p^2)^{- \alpha_F} \label{scalesol}
\ee
The anomalous dimensions $\alpha_G$, $\alpha_F$ fulfill a sum rule,
\be
\alpha_G + 2 \alpha_F = \frac{D - 4}{2} \,. \label{sumrule}
\ee

The approximate DS equations can be solved analytically in the IR regime
\cite{Zwa02,LS02}. There are \emph{two} consistent solutions, 
one of them giving
\be
\alpha_F (D) = \frac{D - 2}{2} \,, \qquad \alpha_G (D) = - \frac{D}{2} \,,
\label{sol1}
\ee
for $2 < D < 4$. The IR exponents of the other solution are obtained, 
for $2 \le D \le 4$, in general numerically from a transcendental equation. 
However, a very good approximation (within 2\% error) is given by
\be
\alpha_F (D) = \frac{D - 1}{5} \,, \qquad 
\alpha_G (D) = - \frac{16 - D}{10} \,. \label{sol2}
\ee
Note that, within the range of dimensions indicated, for both solutions
$\alpha_F > 0$ and $\alpha_G < -1$ hold, which means that the ghost
propagator is enhanced in the IR with respect to the tree-level propagator,
while the gluon propagator tends to zero for $p \to 0$. As a consequence,
in the IR the diagrams with the greatest number of ghost propagators
dominate over the others, a property called ghost dominance. 

Of particular interest is the IR behavior of a dimensionless running
coupling constant defined from the ghost-gluon vertex. Taylor's well-known
argument \cite{Tay71} implies that in the Landau gauge at the symmetrical 
point the proper ghost-gluon vertex receives only finite and scale-independent
quantum corrections, to all perturbative orders. This fact motivates the 
definition
\be
g_R^2 (p^2) = (p^2)^{(D - 4)/2} \, G (p^2) F^2 (p^2) g^2 \label{runncoupl}
\ee
of the dimensionless renormalized coupling constant $g_R$ in terms of the
bare coupling constant $g$. As a consequence of the sum rule \eqref{sumrule},
for the scaling solutions $g_R$ tends toward a nonzero constant in the limit 
$p \to 0$. The value of $g_R(0)$ can be obtained from the IR solutions of the 
DS equations. 

Several years after the discovery of the scaling solutions, however, another
type of solutions of the same approximate DS equations have been found,
called decoupling solutions \cite{AN04,BBL06,Fra08}. 
The latter solutions are characterized by
$\alpha_F = 0$ and $\alpha_G = -1$, for any dimension $D$ in the range
$2 < D \le 4$. In particular, for these solutions the sum rule \eqref{sumrule}
does not hold. Qualitatively, in the decoupling solutions the ghost
propagator is enhanced in the IR at most by a finite factor, while the
gluon propagator tends towards a nonzero constant for $p \to 0$, similar
to the propagator in a massive theory. Consequently, the dimensionless
running coupling constant defined in Eq.\ \eqref{runncoupl} vanishes in the 
IR limit, contrary to the scaling solutions.

Ever since the scaling solutions have been found, lattice calculations in
the Landau gauge have focused on the extreme IR regime. The subject has
been controversial over many years, but as for today there is quite 
convincing evidence that these numerical simulations find gluon and ghost
propagators of the decoupling type in dimensions $D = 3, 4$ 
\cite{BIM07,CM07,SSL07} and scaling solutions in $D = 2$ \cite{Maa07}.

In the present paper, we show that a renormalization group (RG) analysis
in an epsilon expansion is capable of quantitatively reproducing both the 
scaling and the decoupling solutions in the extreme IR,
and furthermore of determining the IR-stability of the solutions. 
Only solutions corresponding to IR-stable fixed points are expected
to be physically realized. Of course, it is well known that in 
the standard formulation of the theory the only perturbatively accesible
fixed point is the trivial one. It is UV-attractive and IR-repulsive and
can therefore not be used for a description of the IR physics. We here
introduce a minimal modification of standard Yang-Mills theory by adding a 
mass term for the gluons as in Ref.\ \cite{CF76}, see also \cite{TW10}. 
The gluon mass is generically expected to be of the order of the QCD 
scale $\Lambda_{\text{QCD}}$. It is well known that this modification
does not affect the UV behavior of the theory. In particular, the beta 
function in this aymptotic region is unchanged and asymptotic freedom is
maintained. A gluon mass term has been considered in many contexts before, 
often on phenomenological grounds. 

Our main motivation for the introduction of a gluon mass term, however,
comes from new theoretical results: the quantization of Yang-Mills theory
in the Landau gauge is complicated by the existence of gauge copies,
and the corresponding restriction of the functional integral over the gauge 
fields to the first Gribov region breaks the BRST invariance of the 
effective action \cite{Zwa93,DGS08}. From the RG viewpoint, the appearance
of a gluon mass term is a natural consequence of the breaking of BRST
invariance. Indeed, it has been shown in Ref.\ \cite{DGS08} that a gluon
mass term arises as a result of the formation of condensates of auxiliary
fields that are introduced in order to implement the restriction
to the Gribov region by adding local operators to the action.
In the present analysis, we will not take any additional auxiliary fields
into account.

In the following, we will hence consider the $D$-dimensional Euclidean action
\be
S = \int d^D x \left( \frac{1}{4} \, F_{\mu\nu}^a F_{\mu\nu}^a +
\frac{1}{2} \, A_\mu^a \, m^2 A_\mu^a +
\d_\mu \bar{c}^a D_\mu^{ab} c^b + i B^a \d_\mu A_\mu^a \right) \label{action}
\ee
with the gluon mass $m$, as the classical action for $SU(N)$ Yang-Mills
theory in the Landau gauge. The Nakanishi-Lautrup field $B^a$ is used to
conveniently implement the restriction to transverse gauge fields,
$\d_\mu A_\mu^a = 0$. As we shall now argue, the existence of a gluon mass 
term profoundly changes the RG analysis of the theory in the
IR. 

Naively, for momenta $p$ with $p^2 \ll m^2$, the mass term in Eq.\ 
\eqref{action} dominates the contributions to the action that are quadratic 
in the gluon field. In fact, this is also true for the loop corrections.
As a relevant example, consider the one-loop contribution to the ghost
self-energy in dimensions $2 \le D \le 4$: it is straightforward to show 
that one may replace the tree-level gluon propagator corresponding to the
action \eqref{action} inside the loop diagram with
\be
\frac{1}{m^2} \left( \delta_{\mu \nu} - \frac{p_\mu p_\nu}{p^2} \right)
\delta^{ab} (2 \pi)^D \delta(p - q) \label{IRprop}
\ee
and obtain exactly the same result in the IR limit, with corrections 
suppressed by powers of $p^2/m^2$, \emph{after renormalization} with an 
appropriate normalization condition for the term $\propto p^2$ in the ghost 
two-point function. For the description of the extreme IR regime, we will 
hence in the following reduce the complete term quadratic in the gluon 
field in the action \eqref{action} to the mass term. 

We initiate the RG analysis by considering the limit $g \to 0$ of the
approximate IR-form of the action which then reduces to
\be
S^0_{\text{IR}} = \int d^D x \left( \frac{1}{2} \, A_\mu^a \, m^2 A_\mu^a +
\d_\mu \bar{c}^a \d_\mu c^a + i B^a \d_\mu A_\mu^a \right) \,.
\label{g0action}
\ee
Under a rescaling of the space-time coordinate $x \to x/s$ with $s > 1$
[to be understood as one step in a Wilsonian RG
transformation for the free field theory defined by the action
\eqref{g0action}], the IR-action is invariant provided that the fields
transform as
\bat
A_\mu^a (x) &\to s^{D/2} A_\mu^a (s x) \,,
&\qquad B^a (x) &\to s^{(D/2) - 1} \, B^a (s x) \,, \n \\
c^a (x) &\to s^{(D/2) - 1} \, c^a (s x) \,,
&\qquad \bar{c}^a (x) &\to s^{(D/2) - 1} \, \bar{c}^a (s x) \label{scaledim}
\eat
($m^2$ being constant under the rescaling). Note that Eq.\ \eqref{scaledim}
now defines the canonical or scaling (mass) dimensions of the fields which
differ, in the case of the gluon field and the Nakanishi-Lautrup field,
from the usual canonical dimensions which refer to the Gaussian fixed point.
To reconcile the scaling dimension of $A_\mu^a$ with the usual canonical
dimension, we may absorb one power of the mass in the gluon field to define
\be
\tilde{A}^a_\mu (x) = m A_\mu^a (x) \label{Atilde}
\ee
and consider $\tilde{A}^a_\mu$ as the fundamental field. Similarly, we define
$\tilde{B}^a (x) = B^a (x)/m$. As far as the gluon field is concerned,
the fixed point action \eqref{g0action} corresponds to the high-temperature
fixed point rather than the usual Gaussian fixed point \cite{Bel91}. 
It is natural to associate the high-temperature fixed point with
a stochastic 
vacuum state.

We will now reintroduce the coupling terms appearing in the Yang-Mills action.
An important consequence of the scaling dimensions \eqref{scaledim} is that
the ghost-gluon, three-gluon, and four-gluon couplings all receive different
scaling dimensions:
\be
g_{\bar{c}Ac} \to s^{1 - (D/2)} \, g_{\bar{c}Ac} \,,
\qquad g_{A^3} \to s^{-1 - (D/2)} \, g_{A^3} \,,
\qquad g_{A^4} \to s^{-D} g_{A^4} \,, \label{coupldim}
\ee
so that all the coupling constants are perturbatively irrelevant [around the 
IR fixed point described by the action \eqref{g0action}], except for
$g_{\bar{c}Ac}$ at dimensions $D \le 2$. In other words, $D=2$ is the upper
critical dimension of the theory, in agreement with an analysis of the 
IR behavior of higher orders in perturbation theory \cite{TW10}. 
Note that the scaling
dimensions \eqref{coupldim} are simply the mass dimensions of the coupling
constants when one replaces the gluon field with the rescaled field
$\tilde{A}_\mu^a$ according to Eq.\ \eqref{Atilde}.

The theory we will hence be considering in the following is defined by the
free action \eqref{g0action} plus the sole interaction term
\be
S^1_{\text{IR}} = \tilde{g} f^{abc} \int d^D x \, \d_\mu \bar{c}^a 
\tilde{A}_\mu^b \, c^c \,, \label{interact}
\ee
where we have redefined the coupling constant as 
$\tilde{g} = g_{\bar{c}Ac}/m$.
It will be interesting to consider this specific theory for dimensions
$D \ge 2$, even though the only IR-stable fixed point for $D > 2$ will turn 
out to be the trivial one \eqref{g0action}, as expected. We
will employ an epsilon expansion around $D=2$, i.e., we will consider
dimensions $D = 2 + \epsilon$.

Note that neglecting the three- and four-gluon couplings (with the argument
that these couplings are perturbatively irrelevant for any positive dimension)
precisely corresponds to ghost dominance because then only the diagrams with
the greatest number of ghost propagators remain. Ghost dominance can hence
be considered as a consequence of the change in the scaling dimension of the
gluon field which, in turn, is a consequence of the appearance of a gluon
mass term.

For the RG analysis of the interacting theory
we calculate the one-loop self-energies, first for the ghost fields.
From the action \eqref{g0action} plus the interaction term \eqref{interact}
one obtains (suppressing the color factor)
\be
 \parbox{3.1cm}{\begin{center}
  \begin{tikzpicture}[>=stealth,scale=1.4]
   \begin{scope}[snake=coil, segment length=3pt, segment amplitude=2pt]
   \foreach \x in {-5,5,...,165}
     \draw[snake] [xshift=0.5cm] (\x+19:0.5cm) -- (\x:0.5cm);
   \end{scope}
   \draw[dash pattern=on 1.5pt off 1.2pt] (-0.4,0) -- (1.4,0) %
node[below] {$\scs p$};
   \draw[->,very thin] (1.18,0) -- (1.15,0);
   \draw[->,very thin] (0.51,0) -- (0.48,0);
   \draw[->,very thin] (-0.22,0) -- (-0.25,0);
   \draw[very thin] (0.5,0.5) -- (0.5,0.5);
   \fill (0,0) circle (1.2pt);
   \fill (1,0) circle (1.2pt);
  \end{tikzpicture}
  \end{center}} 
= - \frac{1}{2} \, \frac{N \bar{g}^2}{4 \pi} \, p^2
\left( \frac{2}{\epsilon} + \gamma_E - \ln (4 \pi)
+ \ln \frac{p^2}{\kappa^2} \right) \label{ghosteps}
\ee
at $D = 2 + \epsilon$, where we have introduced a dimensionless coupling
constant $\bar{g}$ via
\be
\tilde{g}^2 = (\kappa^2)^{-\epsilon/2} \, \bar{g}^2
\ee
[cf.\ Eq.\ \eqref{coupldim}]. For the gluon self-energy, on the other hand,
one obtains in $D = 2 + \epsilon$ dimensions,
\be
\parbox{3.1cm}{\begin{center}
  \begin{tikzpicture}[>=stealth,scale=1.4]
   \begin{scope}[snake=coil, segment length=3pt, segment amplitude=2pt,%
   line before snake=1.5pt]
   \draw[snake] (-0.45,0) -- (0.05,0);
   \draw[snake] (0.98,0) -- (1.44,0) node[below] {$\scs p$};
   \end{scope}
   \draw[dash pattern=on 1.5pt off 1.2pt] (0.5,0) circle (0.5cm);
   \draw[->,very thin] (0.49,-0.5) -- (0.52,-0.5);
   \draw[->,very thin] (0.51,0.5) -- (0.48,0.5);
   \fill (0,0) circle (1.2pt);
   \fill (1,0) circle (1.2pt);
  \end{tikzpicture}
  \end{center}}
= \frac{1}{2} \, \frac{N \bar{g}^2}{4 \pi} 
\left[ \left( \frac{2}{\epsilon} + \gamma_E - \ln (4 \pi)
+ \ln \frac{p^2}{\kappa^2} - 2 \right) \delta_{\mu \nu} 
+ 2 \, \frac{p_\mu p_\nu}{p^2} \right] \,. \label{gluonSE}
\ee
Note that the gluon self-energy contains a longitudinal contribution which
makes the breaking of BRST invariance as a result of ghost 
dominance explicit. In the following we only
consider the transverse part of the self-energy (since the gluon propagator
stays transverse) which is given by the coefficient of $\delta_{\mu \nu}$
in Eq.\ \eqref{gluonSE}. We also emphasize that the counterterm for the
self-energy \eqref{gluonSE} can be chosen \emph{local}.

We now renormalize the theory in the standard way, rescaling the fields
with field renormalization constants $Z_A$ and $Z_c$, and impose the 
following normalization conditions on the propagators of 
the renormalized fields:
\bal
\big\langle \tilde{A}_{R,\rho}^a (p) \tilde{A}_{R,\sigma}^b (-q) 
\big\rangle \, \Big|_{p^2 = \mu^2}
= \frac{1}{Z_A} \big\langle \tilde{A}_\rho^a (p) 
\tilde{A}_\sigma^b (-q) \big\rangle \, \Big|_{p^2 = \mu^2}
&= \delta^{ab} \left( \delta_{\rho \sigma} - \frac{p_\rho p_\sigma}{p^2} 
\right) (2 \pi)^D \delta (p - q) \,, \n \\
\left\langle c_R^a (p) \bar{c}_R^b (-q) \right\rangle 
\Big|_{p^2 = \mu^2}
= \frac{1}{Z_c} \left\langle c^a (p) \bar{c}^b (-q) \right\rangle 
\Big|_{p^2 = \mu^2}
&= \frac{1}{\mu^2} \, \delta^{ab} (2 \pi)^D \delta (p - q) \,. 
\label{normcond}
\eal
Upon substituting the expressions \eqref{ghosteps}, \eqref{gluonSE} for the
self-energies, the renormalization constants become functions of the
renormalization scale $\mu$. The resulting one-loop anomalous dimensions 
read
\bal
\gamma_A = \mu^2 \frac{d}{d \mu^2} \ln Z_A &= \frac{1}{2} \, 
\frac{N \bar{g}^2}{4 \pi} \,, \n \\
\gamma_c = \mu^2 \frac{d}{d \mu^2} \ln Z_c &= - \frac{1}{2} \, 
\frac{N \bar{g}^2}{4 \pi} \,. \label{anomdim1}
\eal
The $\mu$-independence of the bare propagators implies, in general, that
\bal
\mu^2 \frac{d}{d \mu^2} \, \big\langle \tilde{A}_{R,\rho}^a (p) 
\tilde{A}_{R,\sigma}^b (-q) \big\rangle
&= - \gamma_A \, \big\langle \tilde{A}_{R,\rho}^a (p) 
\tilde{A}_{R,\sigma}^b (-q) \big\rangle \,, \n \\
\mu^2 \frac{d}{d \mu^2} \, \left\langle c_R^a (p) 
\bar{c}_R^b (-q) \right\rangle
&= - \gamma_c \left\langle c_R^a (p) \bar{c}_R^b (-q) \right\rangle \,.
\label{ren2ptdiff}
\eal

For the definition of the (dimensionless) renormalized coupling constant
we consider the renormalized proper ghost-gluon vertex at the symmetric point,
\bal
\left. \Gamma_{R, \bar{c} \tilde{A} c} (p, q, r) 
\right|_{p^2 = q^2 = r^2 =\mu^2} 
&= Z_A^{1/2} (\mu) Z_c (\mu) \left. \Gamma_{\bar{c} \tilde{A} c} (p, q, r) 
\right|_{p^2 = q^2 = r^2 =\mu^2} \n \\
&= \mu^{-\epsilon/2} \, \bar{g}_R (\mu) f^{a b c} i p_\rho \, (2 \pi)^D
\delta(p + q + r) \,. \label{gRdef}
\eal
(We have suppressed the Lorentz and color indices on the proper ghost-gluon
vertex $\Gamma_{\bar{c} \tilde{A} c}$ for readability.) Simple power counting 
shows that the loop corrections to $\Gamma_{\bar{c} \tilde{A} c}$ at the
symmetric point are $\mu$-independent (Taylor's theorem). Consequently,
one finds for the beta function to one-loop order,
\be
\beta (\bar{g}_R) = \mu^2 \frac{d}{d \mu^2} \, \bar{g}_R 
= \frac{1}{2} \, \bar{g}_R \left( \frac{\epsilon}{2} - \frac{1}{2} \,
\frac{N \bar{g}_R^2}{4 \pi} \right) \,. \label{beta1}
\ee

For $\epsilon > 0$ ($D > 2$), we find a trivial IR-stable fixed point 
(as expected) and a nontrivial IR-unstable fixed point. Although the latter 
is irrelevant for the IR physics, for the sake of comparison with the
results of the DS analysis it is interesting to determine 
the IR behavior of the propagators corresponding to this fixed point.
Substituting the unstable fixed-point value of $\bar{g}_R$ in Eq.\ 
\eqref{anomdim1} and integrating the differential equations 
\eqref{ren2ptdiff} for the renormalized propagators with these fixed point 
values for the anomalous dimensions and
the normalization conditions \eqref{normcond} as initial conditions yields 
\bal
\big\langle \tilde{A}_{R,\rho}^a (p) \tilde{A}_{R,\sigma}^b (-q) \big\rangle
&= \left( \frac{p^2}{\mu^2} \right)^{\epsilon/2}
\delta^{ab} \left( \delta_{\rho \sigma} - \frac{p_\rho p_\sigma}{p^2} 
\right) (2 \pi)^D \delta (p - q) \,, \n \\
\left\langle c_R^a (p) \bar{c}_R^b (-q) \right\rangle 
&= \frac{1}{p^2} \left( \frac{\mu^2}{p^2} \right)^{\epsilon/2} 
\delta^{ab} (2 \pi)^D \delta (p - q) \,. 
\eal
The IR-limit of the propagators is hence of the form \eqref{scalesol},
and the exponents coincide \emph{exactly} with the scaling solution 
\eqref{sol1} of the DS equations found in Refs.\ \cite{Zwa02,LS02}.

The other fixed point in Eq.\ \eqref{beta1}, which is IR-stable and
hence physically relevant for the description of the IR regime of the theory
in $D > 2$ dimensions, is trivial in accord with our identification of $D=2$
as the upper critical dimension. Apart from a vanishing 
renormalized coupling constant,
one hence finds the ``classical'' behavior of the propagators in the IR limit,
which means in particular that (the transverse part of) the gluon propagator 
tends toward a non-vanishing constant for zero momentum [cf.\ Eq.\
\eqref{g0action}]. The trivial fixed point then corresponds to the
decoupling solution \cite{AN04,BBL06,Fra08} of the DS equations described in 
the paragraph after Eq.\ \eqref{runncoupl}. The triviality of the fixed point 
justifies the use of perturbation theory for the decoupling solution 
\cite{TW10}.

We can obtain a more precise description of the IR behavior of the propagators
and the renormalized coupling constant by analyzing the approach to the 
(stable) IR fixed point as the momentum scale tends to zero. Namely,
integrating the differential equation \eqref{beta1} for $\bar{g}_R (\mu)$
gives
\be
\frac{N \bar{g}_R^2 (\mu)}{4 \pi} = \frac{(\mu^2/\Lambda^2)^{\epsilon/2}}
{1 + (\mu^2/\Lambda^2)^{\epsilon/2}} \, \epsilon \,, \label{decouprunning}
\ee
where $\Lambda$ is the (undetermined) scale with
\be
\frac{N \bar{g}_R^2 (\Lambda)}{4 \pi} = \frac{\epsilon}{2} \,,
\ee
``halfway'' between the two fixed points. Substituting the expression for
$\bar{g}_R (\mu)$ from Eq.\ \eqref{decouprunning} for $\bar{g}$ in the 
anomalous dimensions \eqref{anomdim1} and using the result in
the integration of the differential equations \eqref{ren2ptdiff} for
the renormalized propagators [together with Eq.\ \eqref{normcond}] leads to
\bal
\big\langle \tilde{A}_{R,\rho}^a (p) \tilde{A}_{R,\sigma}^b (-q) \big\rangle
&= \frac{1 + (p^2/\Lambda^2)^{\epsilon/2}}{1 + (\mu^2/\Lambda^2)^{\epsilon/2}} 
\, \delta^{ab} \left( \delta_{\rho \sigma} - \frac{p_\rho p_\sigma}{p^2} 
\right) (2 \pi)^D \delta (p - q) \,, \n \\
\left\langle c_R^a (p) \bar{c}_R^b (-q) \right\rangle 
&= \frac{1}{p^2} \, \frac{1 + (\mu^2/\Lambda^2)^{\epsilon/2}}
{1 + (p^2/\Lambda^2)^{\epsilon/2}} \, 
\delta^{ab} (2 \pi)^D \delta (p - q) \,. \label{decoupl}
\eal
Also note that, from Eq.\ \eqref{runncoupl}, $g_R^2 (p^2) = \bar{g}_R^2 
(\mu^2 = p^2)$ (except for a constant factor from the finite renormalization 
of the ghost-gluon vertex).

The result \eqref{decoupl} for the propagators is at least qualitatively
confirmed by calculations on very large lattices \cite{BIM07,CM07,SSL07}. 
In dimension $D=3$, the linear rise of the gluon propagator with $|p|$ from 
a constant finite value in the IR is clearly seen \cite{CM09} (incidentally, 
this is very different from the usual behavior of a massive propagator). 
In $D=4$ dimensions, a (approximately) quadratic rise of the gluon propagator
with $|p|$ in the IR as predicted by Eq.\ \eqref{decoupl} is found in lattice
simulations with lattice parameter $\beta = 0$, i.e., without any contribution
from the original Yang-Mills action \cite{CM10}. In order to compare
our results in four dimensions to the case of finite $\beta$, we have to
reintroduce the term $(A^a_\mu \, p^2 A^a_\mu)$ in the action, since the latter
term is of the same order in $(p^2/m^2)$ as the correction to the ``classical''
behavior generated by the application of the RG (at $\epsilon = 2$). Given 
that there are no loop corrections to the $(A^a_\mu \, p^2 A^a_\mu)$-term
at one-loop order, the resulting gluon propagator is precisely of the form 
obtained in the refined Gribov-Zwanziger scenario \cite{DGS08}. 

We also mention that
a semi-quantitative reproduction of the lattice results (for the decoupling
solution) over the whole momentum range was achieved in $D=3$ and $4$ 
dimensions in one of the renormalization schemes proposed in Ref.\ 
\cite{TW10}. The scheme in question implements normalization conditions 
different from \eqref{normcond} for the theory defined by the full action 
\eqref{action}. No epsilon expansion is employed in this approach, and the 
only IR fixed point of the coupling constant found in four dimensions is stable
and trivial, while the three-dimensional IR-stable fixed point is nontrivial.

In summary, our RG analysis recovers, in dimensions
$D > 2$, the decoupling solution and one of the scaling solutions
[given in Eq.\ \eqref{sol1}] of the DS equations. Moreover, unlike the
DS equations, the RG analysis is able to show that the
decoupling solution is IR-stable and hence physically relevant, while the
scaling solution \eqref{sol1} is IR-unstable. In dimension $D = 2$,
the beta function \eqref{beta1} is negative for
$\bar{g}_R \neq 0$, hence the only fixed point, the trivial one corresponding
to the decoupling solution, becomes IR-unstable. This is consistent with 
lattice calculations which do \emph{not} find decoupling (but rather scaling) 
behavior in the IR at $D = 2$ \cite{Maa07}.

What about the other scaling solution of the DS equations given in Eq.\
\eqref{sol2}? As we shall now demonstrate, it can also be obtained
from the RG analysis if the so-called horizon condition \cite{Zwa93} of 
the original (not refined) Gribov-Zwanziger scenario,
\be
F (p^2) \to \infty \quad \text{for} \quad p^2 \to 0 \label{horizon}
\ee
in the notation of Eq.\ \eqref{FGdefs}, is implemented as a 
normalization condition. The horizon condition is fulfilled for the scaling
solutions \eqref{sol1} and \eqref{sol2}, but not for the decoupling
solution. Note that we have obtained the scaling solution \eqref{sol1}
from the RG analysis without ever using the horizon condition
explicitly.

Now, in order to implement Eq.\ \eqref{horizon} classically, we change the
free part \eqref{g0action} of the IR-form of the action to
\be
S^0_{\text{IR}} = \int d^D x \left( \frac{1}{2} \, A_\mu^a \, m^2 A_\mu^a +
\frac{1}{b^2} \, \d_\mu \bar{c}^a (-\d^2) \d_\mu c^a + 
i B^a \d_\mu A_\mu^a \right) \,, \label{g0act2}
\ee
where the parameter $b$ with the dimension of mass is introduced in order
to maintain the usual canonical dimensions of the ghost fields. The form
of the ghost term in Eq.\ \eqref{g0act2} is known as a (isotropic) Lifshitz 
point in statistical physics \cite{HLS75,MC98}.

It is clear from Eq.\ \eqref{g0act2} that the scaling dimensions of the
ghost fields differ from the usual canonical dimensions, namely, the
invariance of the free action under the rescaling $x \to x/s$ implies
the field rescalings
\be
c^a (x) \to s^{(D/2) - 2} \, c^a (sx) \,, \qquad
\bar{c}^a (x) \to s^{(D/2) - 2} \, \bar{c}^a (sx) \,. \label{scaledim2}
\ee
As for the coupling constants, the three- and four-gluon couplings scale
as before in Eq.\ \eqref{coupldim}, since the scaling of the gluon field is
unchanged with respect to Eq.\ \eqref{scaledim}, while for the
ghost-gluon coupling we find 
\be
g_{\bar{c}Ac} \to s^{3 - (D/2)} \, g_{\bar{c}Ac} \,.
\ee
As a consequence, the upper critical dimension of the theory is now $D = 6$.
We rewrite the sole interaction term \eqref{interact} as
\be
S^1_{\text{IR}} = \tilde{g} f^{abc} \int d^D x \, \frac{1}{b^2} \, 
\d_\mu \bar{c}^a \tilde{A}_\mu^b \, c^c \,, \label{inter2}
\ee
where we have defined a new coupling constant $\tilde{g}$ as 
$\tilde{g} = (b^2/m) \, g_{\bar{c}Ac}$.

Following in strict analogy the procedure applied before to the action given 
by Eqs.\ \eqref{g0action} and \eqref{interact} (details will be provided 
elsewhere), we again find two fixed points, only that this time the trivial 
fixed point is IR-unstable and the nontrivial fixed point is IR-stable in 
dimensions $D < 6$. Plugging the value for the stable fixed point
into the anomalous dimensions yields a scaling solution with IR 
exponents [in the notation of Eq.\ \eqref{scalesol}]
\be
\alpha_F (D) = \frac{5 D - 6}{24} \,, \qquad 
\alpha_G (D) = - \frac{18 - D}{12} \,. \label{sol3}
\ee

Comparing the result \eqref{sol3} to (the approximation to) the solution 
\eqref{sol2} of the DS equations obtained in Refs.\ \cite{Zwa02,LS02},
we find that both coincide (exactly) at 
$D = 6$, and the difference between them in the range $2 \le D \le 4$ is 
largest for $D = 2$, where $\alpha_F = 1/5$ (exactly) from the solution of the
DS equation and $\alpha_F = 1/6$ from Eq.\ \eqref{sol3}. However, considering
the usually slow convergence (after Borel resummation) of the epsilon
expansion in applications to statistical physics, the coincidence of Eq.\
\eqref{sol3} with the solution \eqref{sol2} of the DS equations is
surprisingly good.

Together with the previous dicussion, we now have found two different
IR-stable fixed points. Which one is physically realized? The realization of
the fixed point corresponding to the IR exponents \eqref{sol3} depends
on the fine-tuning of the ghost term appearing in the action 
\eqref{g0action}, in order to
fulfill the horizon condition. The latter term is perturbatively
relevant with the scaling dimensions \eqref{scaledim2} of the
ghost fields. Hence, unless there is a physical reason for the fine-tuning
of this term or the imposition of the horizon condition (as there actually 
has formerly been thought to be), the fixed point is unstable with respect
to perturbations of the term in question.

In summary, the RG analysis also reproduces the second scaling solution
\eqref{sol2} upon implementing the horizon condition. However, this
condition is IR-unstable under the RG flow, hence in the end the only IR-stable
fixed point is the one corresponding to the decoupling solution (in
dimensions $D > 2$), in agreement with recent lattice simulations.

\begin{acknowledgments}
The author would like to thank David Dudal, Jos\'e Rodr\'iguez-Quintero, 
and Chris Stephens for valuable discussions, and David Dudal and Chris 
Stephens for a critical reading of the manuscript and helpful comments. 
Support by CIC-UMSNH and Conacyt project CB-2009/131787 is gratefully 
acknowledged.
\end{acknowledgments}

\end{document}